# Colloidal test bed for universal dynamics of phase transitions

Adolfo del Campo
Department of Physics, University of Massachusetts, Boston, MA 02125, USA

Early insight on the critical dynamics of phase transitions arose in a cosmological setting in an effort to understand the origin of structure formation in the early Universe. Kibble pointed out that in a spontaneous symmetry breaking scenario, when a system is driven across a phase transition from a high-symmetry phase to a topologically nontrivial vacuum manifold, causally disconnected regions of the system choose independently the broken symmetry (1,2). These conflicting choices result in the formation of topological defects, such as domain walls in a ferromagnet and vortices in a superfluid, to name a couple of familiar examples.
Soon after, Zurek indicated that signatures of universality in the dynamics of a phase transition could be tested in condensed matter systems, e.g., superfluid Helium (3,4). Further, he improved the estimate for the average size of the domains and predicted a universal power law for the density of topological defects as a function of the rate at which the phase transition is crossed. The combination of these ideas is known as the Kibble-Zurek mechanism (KZM) and has been a lively subject of both theoretical and experimental research during the last decades. The abundant attempts to verify the KZM in the laboratory have however faced a variety of shortcomings and while different aspects of the mechanism have been confirmed, a definite test is still missing (5). A remarkable step forward is reported in this issue by Deutschländer et al., who used colloidal monolayers as a test bed for universal critical dyamics with unprecedented accuracy (6).

In a nutshell, the paradigmatic KZM provides a framework to describe the dynamics across a continuous phase transition. At equilibrium, the correlation length diverges as a universal power law in the thermodynamic limit when the critical point $\lambda_c$ is approached by tuning a control parameter $\lambda$, i.e., $\xi = \xi_0/|\epsilon|^\nu$ with $\epsilon = (\lambda_c - \lambda)/\lambda_c$. The equilibrium relaxation time exhibits a similar power-law behavior, $\tau = \tau_0/|\epsilon|^{z\nu} \propto \xi^z$, that is responsible for the critical-slowing down and the nonadiabatic character of the phase transition dynamics. Above, $\nu$ and $z$ are critical exponents associated with the universality class of the transition. An arbitrary time-dependent modulation of $\lambda = \lambda(t)$ leading to a spontaneous symmetry breaking scenario can be linearized around the critical point $\lambda_c$ so that $\lambda(t) \simeq \lambda_c(1 - t/\tau_Q)$ and $\epsilon(t) \simeq t/\tau_Q$, where $\tau_Q$ is the quench time in which the transition is crossed. As illustrated in Fig. 1(a), KZM exploits the adiabatic-impulse approximation to "chop" the evolution through the phase transition in three sequential stages, where the dynamics is quasi-adiabatic, frozen, and quasi-adiabatic again, as the control parameter takes values $\lambda \gg \lambda_c$, $\lambda \approx \lambda_c$ and

$\lambda \ll \lambda_c$, respectively. The crossover between these stages occurs at the freeze-out time $\hat{t}$, when the equilibrium relaxation time matches the time lagged after crossing the critical point. This characteristic time plays a key role in the KZM and already inherits an imprint of universality, scaling with the quench time as $\hat{t} = (\tau_0 \tau_Q)^{1/(1+\nu z)}$. The beautiful insight behind the KZM is the estimate of the average size of the domains by the value of the equilibrium correlation length at the freeze-out time, i.e., $\hat{\xi} = \xi[\epsilon(\hat{t})]$. As a result, the distance between topological defects increases with the quench time following a universal power law of the form $\hat{\xi} = \xi_0 (\tau_Q/\tau_0)^{\frac{\nu}{(1+\nu z)}}$, which is the main prediction of the KZM. Despite the symmetry $\xi[\epsilon(-\hat{t})] = \xi[\epsilon(\hat{t})]$ within this simplified description, numerical simulations indicate that $+\hat{t}$ plays the dominant role in determining $\hat{\xi}$ (7). In addition, coarsening of domains can already occur in the effectively frozen-out stage for very slow ramps (8).

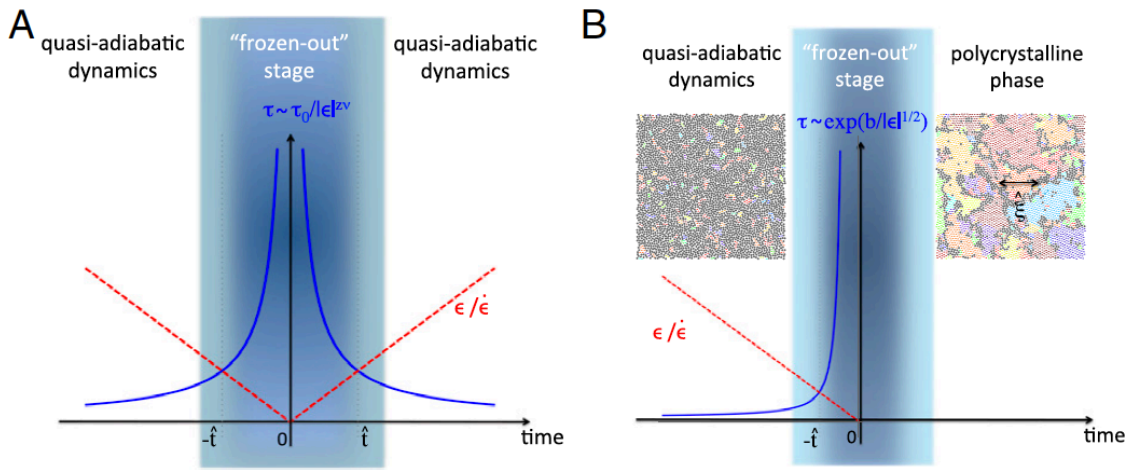

**Fig. 1: Schematic dynamics through a continuous phase transition as described by the Kibble-Zurek mechanism.** (A) In the neighborhood of the critical point, reached at time $t = 0$, the relaxation time diverges and effectively divides the evolution in three sequential stages characterized by a quasi-adiabatic or effectively "frozen-out" dynamics. The equilibrium value of the correlation length at the freeze-out time fixes the size of the domains in the broken symmetry phase. (B) Across a continuous KTHNY phase transition, KZM assumes fluctuations in the isotropic fluid phase to remain frozen during subsequent stages, so that they determine the size of the domains in the resulting polycrystalline phase.

The quest for a conclusive test verifying the KZM scaling faces the following major challenges. Given that the mechanism uses equilibrium properties of the system to account for the nonequilibrium dynamics, measurements of the equilibrium correlation length and relaxation time and the associated critical exponents ($z$ and $\nu$) are required before hand. Important deviations from the power-law behavior are expected in finite-size systems. Both the time-modulation of the control parameter and the system itself should be spatially homogeneous. In addition, the range of available quench rates in the laboratory should be wide enough to test the power-law KZM scaling over several decades, preferably, far away from the onset of adiabatic dynamics. Finally, measurements of the domain sizes should be reliable and performed before standard coarsening and annihilation of topological defects take place, hiding signatures of universality.

Deutschländer et al. study the universal nonequilibrium dynamics induced by cooling at a tunable finite rate a colloidal monolayer. The phase transition under consideration is made clear by the Kosterlitz-Thouless-Halperin-Nelson-Young (KTHNY) theory (9,10,11). At high temperatures an isotropic phase is found with short-range orientational order and isolated disclinations. As the system is cooled down below a critical temperature $T_i$ pairs of dislocation are formed and a hexatic phase emerges with quasi-long-range orientational order. Upon further lowering the temperature below a second critical value $T_m < T_i$, the system enters a crystalline phase characterized by binding of pairs of dislocations. The three phases can be distinguished by a complex order parameter for bond orientation (10). In the crystalline phase, configurations with different global orientations are degenerate. In the course of the evolution, this symmetry is locally broken, resulting in a collage of domains that form a polycrystalline phase. Their average size can be predicted by the KZM using the equilibrium scaling theory above $T_i$, see Fig. 1(b). This assumes that the hexatic phase, being narrow in parameter space, can be ignored and that the dynamics in the crystalline phase is down.

Apart from involving a two-step process, the main peculiarity of the KTHNY universality class for the purpose of testing the KZM is that correlation length and relaxation time do not follow an algebraic divergence at equilibrium, but rather, an exponential one, e.g., $\xi \sim \exp(a/|\epsilon|^{1/2})$ and $\tau \sim \exp(b/|\epsilon|^{1/2})$ with $a, b > 0$ and $\epsilon = (T_i - T)/T_i$. This behavior is predicted by renormalization group (10). The scaling of the correlation length has been experimentally investigated (12). The relaxation time has been considered in theoretical studies of hard disks (13). The exponential equilibrium scaling can be taken into account by suitably modifying the KZM, and defining the freeze-out time accordingly (14,15). Regarding the system size, the number of particles studied is $\sim 5 \times 10^3$, within a sample with over a hundred thousand particles with an extension about hundred times the interparticle distance. These parameters compare favorably to other tests of the KZM (5). Moreover, density gradients in the initial state are suppressed by an exquisite control of the horizontal inclination and the authors assess that temperature gradients are as well absent.

The beads forming the colloid are superparamagnetic and repulsive dipole-dipole interactions can be induced applying an external magnetic field. The control parameter is the ratio of the magnetic energy to the thermal energy and its rate of change can be tuned by nearly three orders of magnitude. The ensuing overdamped nonequilibrium dynamics is analyzed by video microscopy with single-particle resolution that allows unprecedented access to all stages of the phase transition dynamics. The authors record the evolution of the density of defects and the domain size. The characteristic size of the domains is studied as a function of the quench rate and remarkable agreement is found with the KZM prediction $\hat{\xi} = \xi[\epsilon(-\hat{t})]$, when the dynamic critical exponent z is used as a fitting parameter. This agreement suggests that when the quasi-adiabatic dynamics breaks down fluctuations in the initial isotropic fluid freeze in and are preserved across the phase transition, determining the domain size in the resulting polycrystalline phase. Measured data further shows that the domain size can be approximated by a power-law only over a restricted range of quench rates, and with rate-dependent critical exponents (14,15). Yet, the KZM estimate for $\hat{\xi}$ holds even for the slowest cooling ramps where deviations from a power-law behavior in the density of defects are observed. The large fitted value $z = 4.5$ should motivate further analysis.

Overall, the manuscript (6) shows that colloids are an ideal platform to advance our understanding of universal dynamics in critical systems. A variety of exciting prospects for future research can be envisioned. Examples include the critical dynamics of confined colloids across a (zero-temperature) second-order phase transition (16), under inhomogeneous driving (17,18,19), and in the presence of quenched disorder (20).